# Performance Improvement of Graphenic Carbon X-ray Transmission Windows

Sebastian Huebner[1], Natsuki Miyakawa[2], Andreas Pahlke[2] and Franz Kreupl[1]

[1]Department of Hybrid Electronic Systems, Technical University of Munich, Arcisstr. 21, 80333 Munich, Germany
[2]Ketek GmbH, Hofer Str. 3, 81737 Munich, Germany

## ABSTRACT

Graphenic carbon (GC) x-ray transmission windows for EDX and XRF applications with a high transparency for x-rays below 2.5 keV have been fabricated on 6 inch wafers with a CMOS-compatible CVD process. GC windows with an open diameter of 7.4 mm and a thickness of 770 nm withstand up to 6.5 bars of differential pressure. A high transmissivity of 40 % for fluorine K$\alpha$ (0.677 keV) radiation is demonstrated for a GC thickness of 650 nm. The GC membranes outperform beryllium (Be) windows, in terms of higher x-ray transmission and better mechanical stability while avoiding the toxicity of Be. Optical profilometry has been employed to visualize a large deformation of the GC layer during the window fabrication. This seems to limit the thickness of the GC windows that can currently be fabricated. A two-step growth process can overcome these limitations and windows with a thickness of up to 6 μm have been realized.

## INTRODUCTION

X-ray transmission windows need to be highly transparent for the radiation of interest, be helium leak tight, block unwanted stray light and be highly durable. Especially if vacuum encapsulation is required, ambient pressure conditions result in a large mechanical load acting on the transmission window. Graphenic carbon (GC) is a suitable window material that outperforms beryllium (Be) x-ray transmission windows in regard to the x-ray transmission and cycle stability, while avoiding the health concerns of Be [1]. A substrate design incorporating silicon (Si) support bars allows sufficiently thin GC windows that result in an excellent transmission for low energy x-ray radiation with an improved transmission of the carbon K$\alpha$ line compared to available low energy transmission windows based on silicon nitride and polymers [2,3,4].

The required mechanical strength is one of the limiting factors for the transmission of x-ray windows and is the motivation to enhance the mechanical stability of the GC material. In this paper we demonstrate a further improvement of the GC transmission window material regarding the mechanical stability with windows fabricated with a 6 inch compatible CVD process. The mechanical stability of the windows is evaluated by burst pressure testing, cyclic pressure testing and bulge testing. The transmission of the x-ray windows is measured and compared to the alternatives. The GC material exhibits high compressive stress and the window release from the Si substrate is identified as a crucial part due to the deformation of the material stack as the thickness of the Si substrate is gradually reduced. This is, in general, applicable to the fabrication of free standing structures from films with high compressive stress and the observations are therefore described in detail.

## SAMPLE PREPARATION

X-ray transmission windows are fabricated with a 6 inch CVD system using a process described previously [1,5]. Briefly, the procedure is described as follows. 6" silicon wafers, having a thickness of 500 µm or 250 µm, respectively, are used as substrates. A full RCA clean is employed, followed by native oxide removal using hydrofluoric acid with a concentration of 5 %, prior to the GC deposition. A leak test is performed after the substrate is placed in the home-built reactor and the chamber is thoroughly purged with a nitrogen flow of 2000 sccm. The substrate is heated under a high purity hydrogen flow (250 sccm, 99.9999 % purity) until the deposition temperature of 1200°C is reached. High purity methane (2000 sccm, Air Liquide 99.9995 % purity) is used as the precursor gas and introduced to initiate the deposition process. The desired film thickness is obtained by adjusting the growth time, with a growth rate of 40 nm/min. The coated 6 inch wafers are cut into 1 x 1 cm samples for further processing. The GC material exhibits excellent resistance against potassium hydroxide and is therefore a suitable masking material [1]. The GC material covers both sides of the Si-substrate and the backside is structured by oxygen plasma processing to determine the opening geometry of the window [1]. Wet chemical etching (potassium hydroxide with isopropanol as surfactant) is performed to remove the Si substrate in the region where the GC has previously been removed. The freestanding window is formed as the silicon is gradually etched away.

## RESULTS

The maximum pressure stability of the fabricated windows was determined by applying an increasing differential pressure across the window until failure. The maximum pressure was recorded and related to the window thickness, which was measured using a high resolution optical profilometer (Keyence Si-F01) on a pre-structured region of the window or on an adjacent wafer position. Figure 1 (a) shows the maximum pressure stability in relation to the thickness of the window. The fabricated GC windows with an open diameter of approximately 7.2 mm exhibit an improved pressure stability compared to the previously published performance of GC x-ray transmission windows [1]. The earlier observed maximum burst pressure line, indicated as a dashed line in Figure 1 (a), is clearly exceeded.

The observed pressure stability in GC membranes allows for a significant GC thickness reduction leading to transmission windows with an improved x-ray transmission in the energy range below 2.5 keV, while exceeding the mechanical stability of beryllium windows, which are specified to withstand a differential pressure of 2000 mbar at a diameter of 7 mm.

High x-ray transmissivity below 2.5 keV is necessary for efficient light element detection including the elements fluorine, oxygen and carbon. The x-ray transmission was evaluated as described in reference [1] for the energy range between 0.1 - 2.5 keV. The results are plotted in Figure 1 (b), which shows the measured x-ray transmission of a GC window with thicknesses of 650 nm and 1 µm, compared to a commercially available beryllium window.

The measured data indicates an absolute transmission of more than 40 % for the fluorine $K_\alpha$ (0.677 keV) radiation for the GC window with a thickness of 650 nm. This allows the efficient detection of fluorine, which was previously not possible with beryllium windows but required other, special low energy transmission windows.

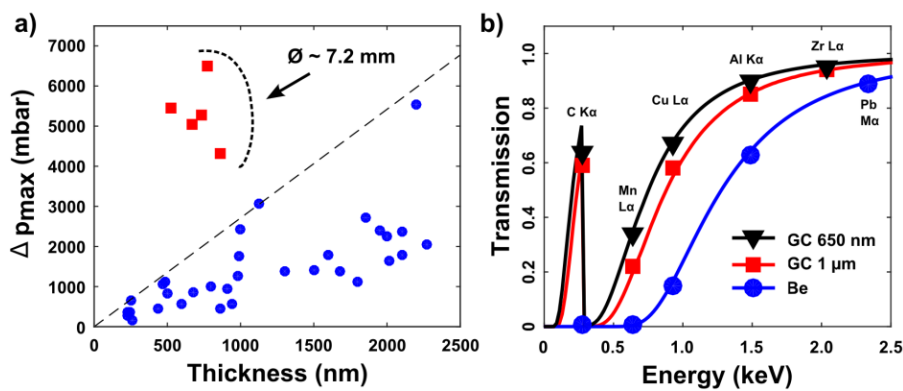

**Figure 1.** a) shows the measured maximum differential pressure required to break a 7 mm wide window for the corresponding GC thickness. The GC windows fabricated on 6 inch wafers (■) are compared to previously published data fabricated with 1x1 cm² substrates (●) [1]. The measured x-ray transmission of GC windows with a thickness of 650 nm and 1 μm, and for a commercial beryllium window for the energy range of 0.1 - 2.5 keV are plotted in (b). The transmission is measured for the indicated discrete energies. The continous data points are fitted to the data by the model of Henke et al. [6].

Pressure cycle testing was performed by applying a differential pressure across the window with a frequency of 3 Hz and a minimum differential pressure of 2400 mbar per cycle. The maximum and minimum pressure values were recorded for each cycle. The GC window with a thickness of 650 nm and a diameter of 7.21 mm exhibits an excellent mechanical stability which is demonstrated in Figure 2. A differential pressure test up to 4000 mbar, as displayed in Fig. 2 (a), did not lead to window failure and helium leak tests confirmed the integrity of the window even after more than 2 million cycles, shown in Fig. 2 (b) with a differential pressure of at least 2400 mbar per cycle.

Bulge testing was performed in order to approximate the Young's modulus of the window material. The procedure and limitations of this method are discussed in detail in reference [2]. Figure 3 shows the results of bulge testing performed with a GC window with a thickness of 710 nm and a diameter of 7.31 mm. The differential pressure induced a measured window deflection, shown in Figure 3 (a), that allows the extraction of the strain and stress of the GC window material, which is plotted in Figure 3 (b). The strain and stress relationship is used to approximate the Young's modulus of the material to 170 GPa, assuming 0.16 as the poissons ratio, which is the value of the basal plane in bulk graphite [7]. The compressive stress of the window material leads to wrinkle formation and the coresponding slack of the window hinders the exact characterisation of the material using the bulge testing method.

Wafer bow measurements (not shown) indicate a high compressive stress in the GC film, which leads to a significant deformation of the Si substrate. It was observed that the fabrication of x-ray windows with a window thickness above a certain threshold resulted in damaged windows with large tears in the GC material. These defects occurred close to the edge of the window geometry. An optical profilometer (Keyence Si-F01) with a high vertical resolution of 25 nm and a motorized stage with an x-y resolution of 200 µm was used to probe the deformation of the sample during the window release. The wet etching process was identified as a possible source of the observed defects.

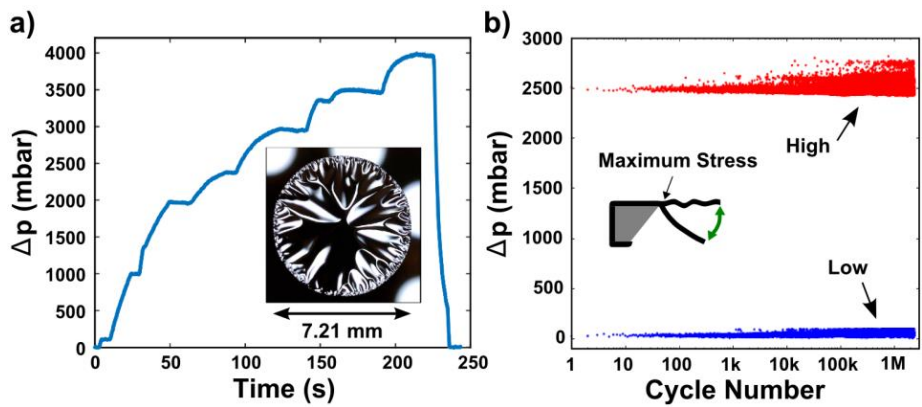

**Figure 2.** a) shows the increasing differential pressure versus time that is applied to a GC window with a thickness of 650 nm and a diameter of 7.21 mm. The differential pressure is gradually increased and the pressure valve closed at given intervals to verify the integrity of the window. The membrane did not break and at 230 seconds, the applied pressure is relieved. The inset shows the tested window with the characteristic wrinkles that form due to the compressive stress of the GC material if no differential pressure is applied. The data of the subsequent pressure cycle test are presented in (b). A differential pressure of at least 2400 mbar is applied which results in a large stress at the window edge. A pressure reservoir with a pressure of 3000 mbar versus atmosphere is used which results in infrequent overshoots of the applied differential pressure.

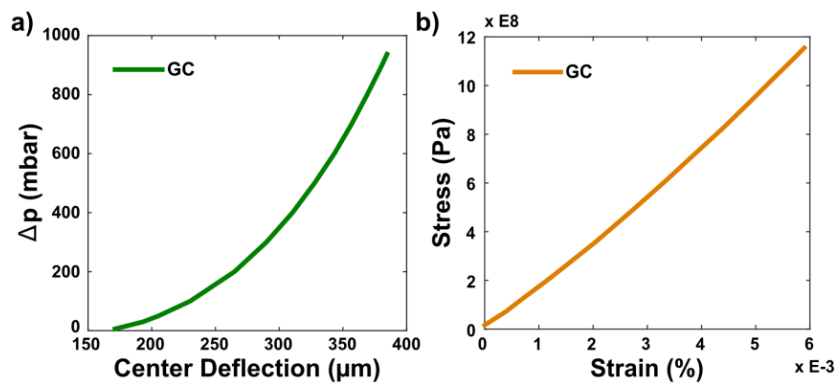

**Figure 3.** a) shows the measured deflection of the window center as an increasing differential pressure is applied across a GC window with a window diameter of 7.31 mm and GC thickness of 710 nm during bulge tests. The extracted strain and stress dependency for the window is plotted in b).

Figure 4 (a) shows the measurement setup and sketches the consequences that arise from a gradually reduced Si substrate thickness: The stress in the GC layer leads to a large deformation of the surface and induces etch inhomogeneities in the silicon.

If we consider the Stoney equation that is commonly used for extracting the material stress from wafer bow measurements, it is apparent that the observed bow is strongly dependent on the thickness of the substrate (Equation 1) [8]. With K denoting the wafer curvature, $E_s$ the biaxial modulus of the substrate, $\sigma_c$ the stress of the coating layer and $t_c$ and $t_s$ the thickness of the coating layer and the Si substrate, respectively the wafer curvature is given by:

$$K = \frac{6t_c}{E_s t_s^2} \sigma_c. \tag{1}$$

During the window release, the remaining silicon thickness is reduced to zero which leads to an increasing curvature, and thus deformation, of the remaining GC-silicon stack. Figure 4 (b) shows the deformation in an x-y scan across a window sample, prior to a complete removal of the Si substrate. The probed sample had a window diameter of 5.8 mm and a GC thickness of 1.2 µm. The 3D scan was performed after 5 hours and 30 minutes of wet etching and the thickness of the Si substrate was therefore reduced to 28 µm at the center of the window.

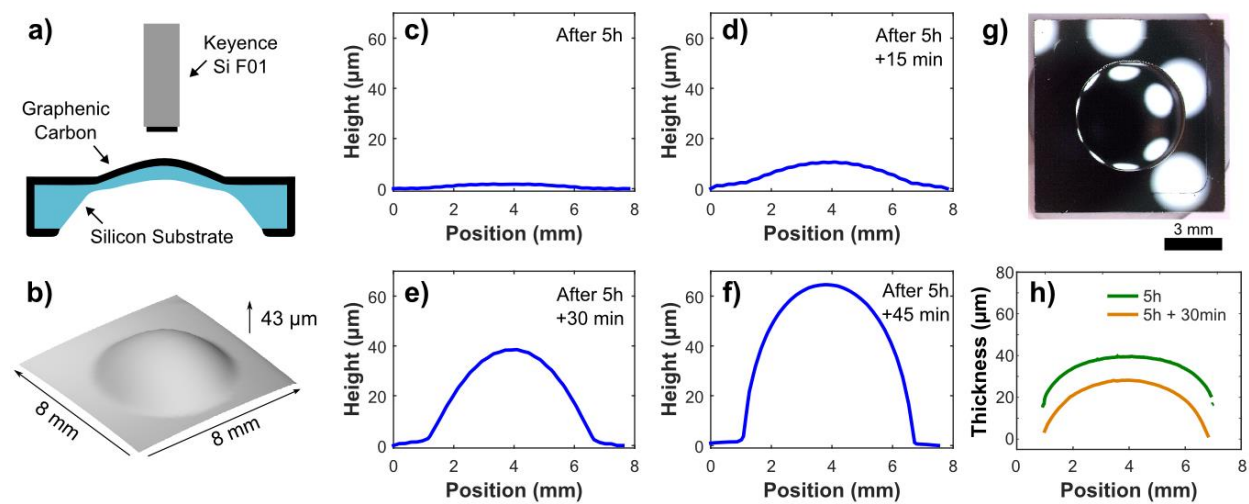

**Figure 4.** a) shows the measurements setup using an optical profilometer to evaluate the window deformation during wet etching of the Si substrate. The sensor is at a fixed position while the sample is positioned using an x-y stage (not shown). b) displays the 3D data obtained with a window prior to the complete removal of the Si substrate in the window region. c)-f) show the deformation of a window as the silicon is gradually removed from the window region. The profiles across the window were extracted from the obtained 3D data. The thickness of the remaining Si substrate corresponds to 43 µm (c), 37 µm (d), 28 µm (e) and finally to a thickness of 18 µm (f) which leads to a growing deformation of the window. An optical image of the examined sample corresponding to the profile shown in f) is depicted in g). The remaining silicon thickness that leads to the sample deformation was also measured in h) and reveals inhomogeneous etching of the silicon.

Two-dimensional profilometer scans were performed on the sample window during the wet etching process and line profiles across the window extracted for measurements after 5 hours (Figure 4 (c)), after 5 hours and 15 minutes (Figure 4 (d)), after 5 hours and 30 minutes (Figure 4 (e)), and after 5 hours and 45 minutes (Figure 4 (f)) of wet chemical etching. In the progress, the remaining Si substrate thickness, at the window center, was successively reduced to 43 µm, 37 µm, 28 µm and finally to a thickness of 18 µm. The large deformation leads to a bending angle approaching 90° at the very edge of the etched region, as it is especially visible in Figure 4 (f). At this point, the deformation is clearly visible with the naked eye as shown in Figure 4 (g). Measurements performed with a dual sensor head setup that is able to measure the remaining Si substrate thickness, shows that the thickness distribution is not uniform but with a minimum thickness near the edge of the window (Figure 4 (h)). This implies that the silicon is removed at first at the very edge of the GC window where the curvature is the largest and where the defects were observed. The defects are asumed to arise only if the resulting bending force is above a certain threshold which is dependent on the thickness and the stress of the GC layer and explains why the defects are only observed on thicker GC windows. In order to overcome this problem we established a two-step approach. Firstly, windows with a small thickness are fabricated that do not suffer from the observed implications during the window release. Secondly, a subsequent growth process is performed with the fabricated GC windows replacing the substrate until the desired window thickness is reached. Using this approach we could demonstrate windows with a thickness of up to 6 µm without the previously witnessed defects.

## SUMMARY

The fabrication of free standing membranes from materials with a high compressive stress using anisotropic wet etching can lead to tears and defects in the resulting membrane due to the large deformation prior to the membrane release. A proposed two step deposition process overcomes the encountered thickness limitation which opens new applications such as large area GC x-ray windows that require an even higher, mechanical stability.

## ACKNOWLEDGMENTS

This work has been supported financially by the Bavarian Ministry of Economic Affairs and Media Energy and Technology under contract number MST- 1210-0006/BAY 177/001. The authors would like to thank Silke Boche and Lukas Holzbaur for their ongoing contribution and support.